\documentstyle[raulbib,epsf]{mn}
\title{\bf Galaxy Formation and Evolution: Low Surface Brightness Galaxies} 
\author[R. Jimenez et al.]{Raul Jimenez$^1$, Paolo Padoan$^2$, Francesca
Matteucci$^3$, Alan F. Heavens$^1$ \\
$^1$Institute for Astronomy, University of Edinburgh, Royal Observatory, 
Blackford Hill, Edinburgh EH9-3HJ, UK. \\
$^2$Instituto Nacional de Astrof\'{\i}sica, \'{O}ptica y Electr\'{o}nica,
Apartado Postal 216, 72000 Puebla, M\'{e}xico. \\
$^3$Department of Astronomy, University of Trieste, SISSA/ISAS, Via 
Beirut 2-4, 34014 Trieste, Italy.}

\begin{document}
\maketitle
\begin{abstract}
We investigate in detail the hypothesis that low surface brightness galaxies
(LSB) differ from ordinary galaxies simply because they form in halos with
large spin parameters.  We compute star formation rates using the Schmidt law,
assuming the same gas infall dependence on surface density as used in models
of the Milky Way.  We build stellar population models, predicting colours,
spectra, and chemical abundances.  We compare our predictions with observed
values of metallicity and colours for LSB galaxies and find excellent
agreement with {\it all} observables. In particular, integrated colours,
colour gradients, surface brightness and metallicity match very well to the
observed values of LSBs for models with ages larger than 7 Gyr and high values
($\lambda > 0.05$) for the spin parameter of the halos.  We also compute the
global star formation rate (SFR) in the Universe due to LSBs and show that it
has a flatter evolution with redshift than the corresponding SFR for normal
discs.  We furthermore compare the evolution in redshift of $[Zn/H]$ for our
models to those observed in Damped Lyman $\alpha$ systems by
\scite{Pettini+97} and show that Damped Lyman $\alpha$ systems abundances are
consistent with the predicted abundances at different radii for LSBs.
Finally, we show how the required late redshift of collapse of the halo may
constrain the power spectrum of fluctuations.
\end{abstract}


\section{Introduction}

Late type low surface brightness galaxies are found to be a significant
fraction of the total galaxy population in the Virgo cluster
(\pcite{Impey+88}), in the Fornax cluster (\pcite{Irwin+90}), and in the field
(\pcite{Mcgaugh+95}; \pcite{Sprayberry+96}; \pcite{Sprayberry+97}).  A theory
of galaxy formation should therefore account for the existence of LSBs.  On
the other hand, LSBs are particularly useful because they are simpler objects
than high surface brightness (HSB) galaxies: they have relatively little
present day star formation and little dust (reddening). Therefore it is easier
to model their stellar population and star formation history.  They are also
more dark matter dominated than HSBs, and therefore particularly suitable for
probing the structure of dark matter halos.  In order to compare theories of
galaxy formation with observational surveys of galaxies it is important not
only to quantify observationally the abundance of LSBs, but also to understand
their nature. For instance, if LSBs are made of very young stellar
populations, as often claimed in the literature, they may be irrelevant in the
picture of the Universe at high redshift, while they could be a substantial
component of that picture if their stellar populations are old.

The existence of LSB galaxies can be readily understood if it is assumed,
following \scite{Fall+80}, that the specific angular momentum of the baryons
is approximately conserved during their dissipation into a rotationally
supported disk, and that the disk length--scale is therefore related to the
angular momentum of the dark matter halo. The low surface brightness is the
consequence of the low surface density of the disk, which is due to the larger
spin parameter of the dark matter halos of LSBs relative to the spin parameter
of the halos of HSB disks (e.g. \pcite{Dalcanton_disc+97}; \pcite{Mo+97};
\pcite{Jimenez+Heavens+97}).

However, while the surface density of a disk can be simply related to the mass
and spin parameter of its dark matter halo, its surface brightness, and
therefore its mass to light ratio ($M/L$), depends on the stellar population
and could in principle vary with galaxy type and age. For instance, $M/L$ must
have a significant time dependence due to the history of star formation in the
disk and to the continuous death of massive stars. Moreover, $M/L$ in a given
photometric band changes with time also as a consequence of the colour
evolution of the stellar population, which is sensitive also to the history of
the chemical enrichment of the disk.  Therefore, a comparison between a simple
model of disk formation and the observations cannot be made without an
appropriate model for the stellar population in the disk that includes {\it
self-consistently} the chemical evolution of the population.

\begin{figure}
\centering
\leavevmode
\epsfxsize=1.
\columnwidth
\epsfbox{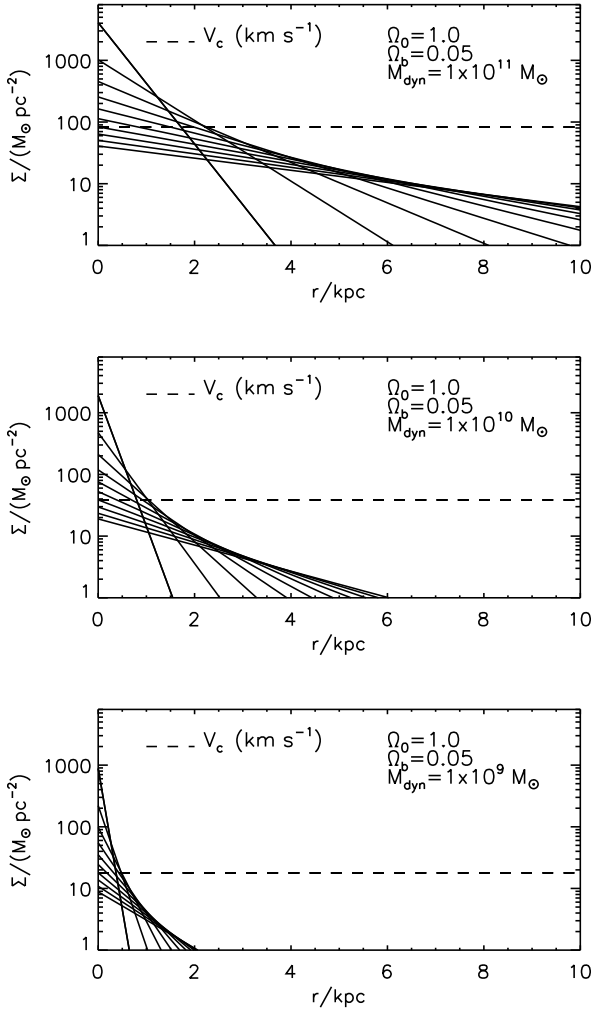}
\caption[]{Initial surface density for an isothermal sphere and 3 
different masses. The solid lines correspond to $\lambda=0.1$ to 0.01 (from 
top to bottom) in steps of 0.01. The circular velocity of the halo (dashed
line) is also plotted.}  
\label{f1}
\end{figure}

In this paper we model self-consistently the chemical and photometric
evolution of the stellar population of LSB disks, formed in dark matter halos
described by isothermal spheres.  The aim of this paper is twofold:

\begin{enumerate}
\item show that only one parameter, namely the {\it spin parameter $\lambda$} 
of the dark halo, can explain the surface brightness of LSBs.
\item use LSB disk galaxies to study the formation and evolution of all disk
galaxies.
\end{enumerate}

The second point is motivated by the first, that is by the fact that LSB disks
are in fact so similar to HSB disks. On the other hand, LSBs are more dark
matter dominated than HSBs, and therefore better described by a very simple
model where only the gravitational field of an isothermal halo is considered.

Our assumptions are as follows:

\begin{enumerate}
\item{The specific angular momentum of baryonic matter is the same as for 
dark matter}
\item{Gas settles until centrifugally supported, in a given dark matter halo}
\item{The star formation rate is given by the Schmidt law}
\item{The gas infall rate is assumed to be the same function of total 
surface density as used in models of the Milky Way}
\end{enumerate}

\begin{figure*}
\centering
\leavevmode
\epsfxsize=1.6
\columnwidth
\epsfbox{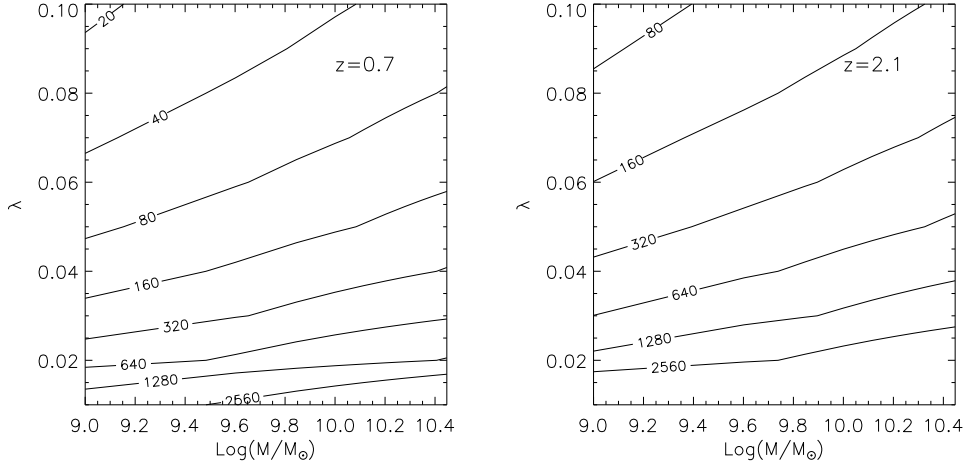}
\caption[]{Contour plot of the value of the central surface density 
(in M$_{\odot}$pc$^{-2}$) for different values of the halo mass and 
spin parameter. Left panel: halo formation redshift $z=0.7$. Right 
panel: halo formation redshift $z=2.1$.}
\label{f2}
\end{figure*}

We first use the density profile of the halo to compute the surface density of
the settling disk. We then use the Schmidt formation law and an infall rate
that reproduces the observed properties of the Galaxy in conjunction with the
chemical evolution models by \scite{Matteucci+89} to compute the star
formation rate at several radii of the disk and the evolution of several
chemical species (H, D, He, C, N, O, Ne, Mg, Si, S, Ca, Fe and Zn). We then
proceed to compute spectra, integrated colours, colour profiles and surface
brightness for two different values of the spin parameter of the halo.

We first present in the next section the simple non--self--gravitational disk
model and the non--singular halo model. We describe in section 3 the set of
synthetic stellar population models used to predict the spectra and colours of
the stellar population.  In section 4 we model the star formation in the disk
with a Schmidt law and a gas infall law, assuming the same model of the
chemical evolution of the Galaxy as \scite{Matteucci+89}. Finally, in section
5 we compute synthetic spectra and photometric properties of the stellar
population in the disk, using the stellar evolution tracks of JMSTAR15,
stellar atmosphere models by \scite{Kurucz_92} and J{\o}rgensen (private
communication) and the chemical evolution models built for LSBs (see section
4).  We discuss the results in section 6.

The most important results of this work are:

\begin{enumerate}
\item Observed colour profiles, chemical abundances and surface brightness
profiles for LSBs are well fitted if they are assumed to have
a spin parameter for its halo higher than HSBs.
\item LSBs are not young objects, as often claimed in the literature, since
their colours are well fitted by old ($> 7$ Gyr) stellar populations.
\item There is discrepancy between the photometric age of the galaxies and 
the age of formation of their halo, which indicates that the star formation 
can start about 2~Gyr before the halo is formed. This discrepancy is reduced 
to 1~Gyr if the Hubble constant is assumed to be $H_0=65$~kms$^{-1}$Mpc$^{-1}$
and completely removed if the Universe is open ($\Omega < 0.3$) or has a
significant vacuum energy contribution ($\Lambda > 0.6$).
\item The earliest stellar populations of LSBs could be present in the high
redshift universe in a similar proportion to HSBs as they are now. Their
colours at high redshift (as the colours of all disks) are about 1~mag bluer
than at low redshift.
\end{enumerate}

In a previous paper (\pcite{Padoan+Jimenez+Antonuccio97a}) we have shown that
LSBs are not necessarily young and un-evolved systems (e.g.
\pcite{McGaugh+Bothun94,deBlok_phot+95}).  We used the simplest possible
model, a burst of star formation, to determine a lower limit to the age of
LSBs studied by \scite{deBlok_phot+95}. In the present work we considerably
improve our previous model because we use a more realistic continuous star
formation process, we include self--consistently the chemical evolution, and
we connect the disk model to the cosmological scenario using the spherical
collapse model (\pcite{Gunn_Gott_72}).  This more detailed model confirms our
previous result that the blue LSB disks in the sample by
\scite{deBlok_phot+95} are not un-evolved objects collapsed at late times from
low initial over-densities (\pcite{McGaugh+Bothun94,Mo+94}), but rather
normally evolved disk galaxies.

Throughout the paper the present day value of the Hubble constant is 
assumed to be $H_0=75$~kms$^{-1}$Mpc$^{-1}$.

\section{The Halo and Disk Models}

In this work we model LSB galaxies as non--self--gravitating disks inside
isothermal dark matter halos, following \scite{Mo+97}.  Isothermal spheres are
found to present a rough description of halos in numerical simulations
(\pcite{Frenk+85,Frenk+88,Quinn+86,Navarro+97}), and have been previously used
in galaxy formation models by \scite{Kauffmann+93} and \scite{Cole+94}.

We briefly summarize the non--self--gravitating disk model as presented in
\scite{Mo+97}. In the spherical collapse model (Gunn \& Gott 1972), the halo
mass $M$ and its circular velocity $V_c$, at redshift $z$, are:
\begin{equation} 
M=\frac{V_c^3}{10GH(z)}
\label{1}
\end{equation} 
where $H(z)$ is the Hubble constant at redshift $z$:
\begin{equation}
H(z)=H_0[\Omega_{\Lambda,0}+(1-\Omega_{\Lambda,0}-\Omega_0)(1+z)^2
+\Omega_0(1+z)^3]^{1/2}
\label{2}
\end{equation}
where $\Omega_0,\Omega_{\Lambda,0}$ and $H_0$ are the values at $z=0$.
Note that the halo is characterised by a circular velocity $V_c$, but there
is no implication that the halo is rotating.  $V_c$ is the speed required
for centrifugal support.  In general, the halo is not expected to be
rotationally supported.

The disk is assumed to be thin and in centrifugal balance, to have an
exponential surface density profile, and an angular momentum and a mass that
are a fraction $f_b$ of the angular momentum and mass of the dark matter
halo. The value of $f_b$ is here assumed to be equal to the ratio of 
the baryonic to total matter density:
\begin{equation}
f_b=\Omega_{b,0}/\Omega_{0}
\label{3}
\end{equation}
The disk mass $M_d$, surface density $\Sigma_d$, and scale--length $R_d$ are:
\begin{equation}
M_d=\frac{f_bV_c^3}{10GH(z)}=2\pi\Sigma_0R_d^2
\label{4}
\end{equation}
\begin{equation}
\Sigma_d(R)=\Sigma_0\exp(-R/R_d)
\label{5}
\end{equation}
\begin{equation}
R_d=\frac{\lambda V_c}{10 \sqrt{2} H(z)}
\label{6}
\end{equation}
where $\Sigma_0$ is the central surface density
\begin{equation}
\Sigma_0=\frac{10f_bV_cH(z)}{\pi G\lambda^2}
\label{7}
\end{equation}
and $\lambda$ is the spin parameter of the halo:
\begin{equation}
\lambda=J|E|^{1/2}G^{-1}M^{-5/2}
\label{8}
\end{equation}
and $E$ is the total energy of the halo.

In Fig.~\ref{f1} surface density profiles are plotted for $z=0.7$ for three
different values of the halo mass. In each plot the profiles for different
values of $\lambda$ are shown, with $0.01\le\lambda\le0.1$. Also shown is the
circular velocity. Models with larger $\lambda$ have lower central surface
density and larger disk scale--length. Fig.~\ref{2} is a contour plot of the
value of the central surface density for different values of the halo mass and
$\lambda$. As can be seen in equation (\ref{7}), $\Sigma_0\propto
M^{1/3}\lambda^{-2}$. A variation in $\lambda$ between $0.04$ and $0.1$
corresponds therefore to a change in $\Sigma_0$ of a factor $6.25$, which is
equivalent to about 2 magnitudes in surface brightness, from the Freeman law
value of $\mu_B(0)=21.6$~mag~arcsec$^{-2}$ (\pcite{Freeman_70}), to
$\mu_B(0)=23.6$~mag~arcsec$^{-2}$, if $M/L$ of the baryons does not depend on
$\lambda$.  On the other hand, the disk scale--length scales like
$R_d\propto\lambda$, so a change in $R_d$ of a factor $2.5$ can be
achieved. Since the halo is isothermal ($\rho\propto R^{-2}$), its mass grows
proportionally to the distance from the center, and the ratio of dynamic mass,
$M_{dyn}$, to disk mass, is therefore proportional to the extension of the
disk,
\begin{equation}
M_{dyn}/M_{disk}\propto\lambda
\label{9}
\end{equation}
because more extended disks contain a larger fraction of the dark matter halo
than less extended disks, and the dark halo dominates the gravitational
potential. This explains the fact that LSBs typically have 
$M_{dyn}/L$ about twice as large as HSBs, while they still obey the same
Tully-Fisher relation as HSBs (\pcite{Zwaan+95}), as long as the $M/L$ of the
baryons is not significantly different in the two classes of galaxies:
equations (1) and (6) imply $V_c^3 \propto M_{dyn}/\lambda \propto M_{disk}$.

\subsection{Isothermal Halos and Disk Properties}

In this work we refer to the sample of LSB disk galaxies by
\scite{deBlok_phot+95}, for the photometry, and by \scite{deBlok_HI+96}, for
the HI profiles and rotation curves. The HI data for U0128, U1230, U5005,
U5750, and U6614 are from \scite{VanDerHulst+93}. Photometric data for the
galaxies F563-V2 and F571-8 are taken from \scite{McGaugh+Bothun94} and
\scite{deBlok+McGaugh97} respectively. Only 14 galaxies with HI rotation
curves that clearly flatten have been used for the discussion in this
section. The mean central surface brightness of the 14 galaxies is
$\mu_B(0)=23.5$~mag~arcsec$^{-2}$, the mean disk scale--length $R_d=4.7$~kpc
($H_0=75$~kms$^{-1}$Mpc$^{-1}$), and the mean circular velocity
$V_c=104$~kms$^{-1}$.

\begin{figure}
\centering
\leavevmode
\epsfxsize=1.
\columnwidth
\epsfbox{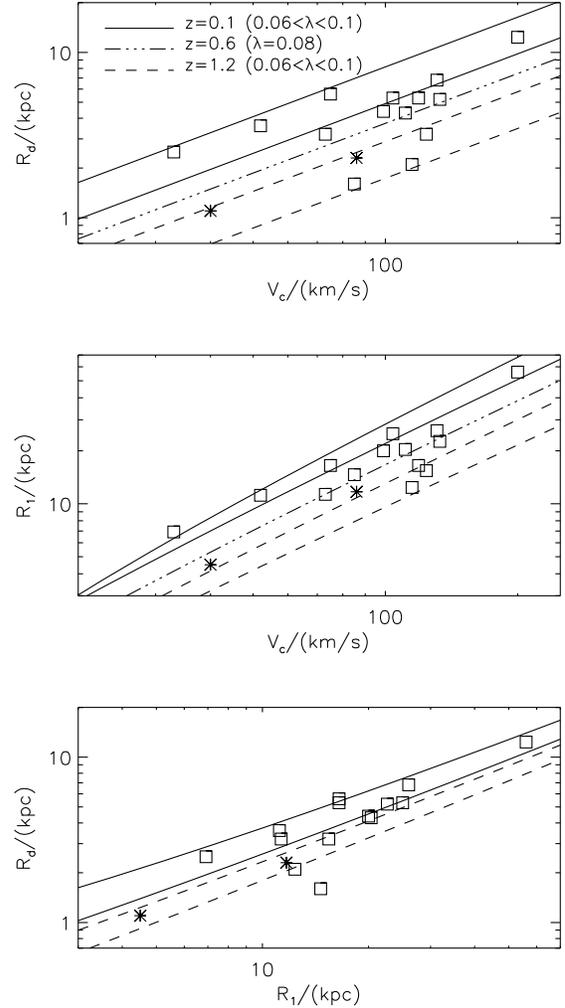}
\caption[]{Upper panel: The HI disk scale--length of the models, $R_d$, versus
the circular velocity, for different values of formation redshifts $z$ and
spin parameter $\lambda$. The squared symbols are 14 LSB galaxies, and the
asterisks the two theoretical models for which the chemical and
spectro--photometric evolutions are calculated in the paper. The scale--length
of the LSB galaxies is the one measured in the B photometric band. The plot
shows that, for a range of $\lambda$ that should be appropriate for LSBs, the
halos of the galaxies in the sample should turn-around in the redshift range
$0.1<z<1.2$.

Middle panel: The radius, $\equiv R_1$, at which the HI surface density is
equal to $1$~M$_{\odot}$pc$^{-2}$, versus the circular velocity, for the
theoretical models and for the galaxies in the sample (squared symbols).  The
asterisks are again the two models discussed in the paper.

Bottom panel: The disk scale--length is plotted versus $R_1$. All symbols are
like in the previous two panels.  The galaxy that appears inconsistent with
the model is F583-1.}
\label{f3}
\end{figure}

In Fig.~\ref{f3} (upper panel) the HI disk scale--length of the models, $R_d$,
is plotted versus the circular velocity, for different values of formation
redshifts $z$ and spin parameter $\lambda$.  Also plotted are the 14 LSB
galaxies (squared symbols) and two theoretical models for which the chemical
and spectro--photometric evolutions are calculated in this work (asterisks).
The scale--length of the galaxies is the one measured in the B photometric
band. The plot shows that, for a range of $\lambda$ that should be appropriate
for LSBs, the halos of the galaxies in the sample should turn-around in the
redshift range $0.1<z<1.2$.

\begin{table}
\begin{center}
\begin{tabular}{lcccc}
&  $Vc/(km s^{-1})$ &  $M/(1.e10 M_{\odot}$) & $\lambda$ &  $z$  \\
\hline\hline
F561-1   &  52   &       3.6   &       0.09 & 0.1 \\
F563-1   & 111   &      26.4   &       0.07 & 0.4 \\
F563-V1  &  30   &       1.2   &       0.07 & 0.0 \\
F563-V2  & 115   &      20.0   &       0.05 & 0.8 \\
F568-1   & 119   &       7.9   &       0.30 & 2.5 \\
F568-V1  & 124   &      16.0   &       0.10 & 1.4 \\
F571-8   & 133   &      26.3   &       0.11 & 1.0 \\
F571-V1  &  73   &       4.4   &       0.13 & 1.0 \\
(F583-1  &  85   &     300.0   &       0.001 & -0.8) \\
U0128    & 131   &      26.6   &       0.14 & 0.9 \\
U1230    & 102   &      39.0   &       0.05 & 0.0 \\
U5005    &  99   &      22.9   &       0.06 & 0.2 \\
U5750    &  75   &       7.5   &       0.14 & 0.4 \\
U6614    & 200   &     178.5   &       0.09 & 0.2 \\
\hline\hline
mean:    & 105   &      29.2   &      0.11 &  0.7 \\
\hline\hline
\end{tabular}
\caption[]{Circular velocity, halo mass, spin parameter and halo
formation redshift from equations (10)-(12), for 14 LSB galaxies 
with rotation curves that clearly flatten. F583-1 has no solution
and is not included to compute the mean values.}
\end{center}
\end{table}

\begin{figure}
\centering
\leavevmode
\epsfxsize=.8
\columnwidth
\epsfbox{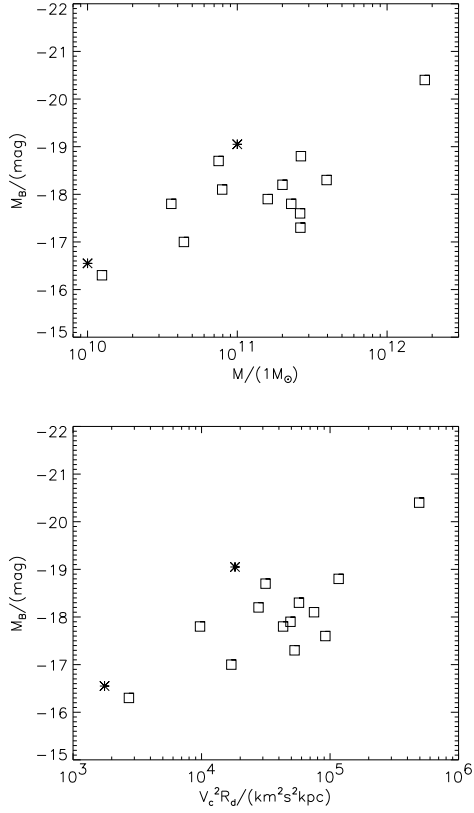}
\caption[]{The absolute blue magnitude of the galaxies is plotted versus
the estimated halo mass (from Table~1) and the value of $V_c^2R_d$, that
should be proportional to the halo mass, contained inside the radius $R_d$. In
both plots the absolute magnitude grows with the mass estimator.}
\label{f4}
\end{figure}

A different way to estimate the size of the galaxies in this sample is to use
the radius, $\equiv R_1$, at which the HI surface density is equal to
$1$~M$_{\odot}$pc$^{-2}$. The value of $R_1$ is plotted versus the circular
velocity in Fig.~\ref{3} (middle panel) for the theoretical models and for the
galaxies in the sample. The two models discussed in this work are also marked
(asterisks). This plot confirms the result of the previous one: the halos of
the LSB galaxies in this sample are formed at a redshift in the range
$0.1<z<1.2$.  Notice that the radius $R_1$ is typically very large compared to
the optical disk of the galaxy, so that virtually no star formation has
occurred at that radius. This is also suggested by the fact that the HI
profiles look very close to exponential around $R_1$, while most of the gas
has clearly been lost into stars at inner radii. Therefore, the value of $R_1$
is suitable for a direct comparison between the model prediction and the
observations. Conversely, the HI profiles are flattish or with holes at inner
radii, which suggests that most of the gas has turned into stars inside $R_d$
and therefore the value of $R_d$ can also be compared directly with model
predictions.

The strong correlation between the value of the B band disk scale--length and
the value of $R_1$, relative to the models, is also shown in the bottom panel
of Fig.~\ref{f3}, where $R_d$ is plotted versus $R_1$. All galaxies but one
are again described by formation redshift in the range $0.1<z<1.2$.  The
galaxy that appears inconsistent is F583-1. This galaxy has an extremely small
scale--length $R_d=1.6$~kpc (\pcite{deBlok+McGaugh97}), given its circular
velocity $V_c=85$~kms$^{-1}$.

\begin{figure}
\centering
\leavevmode
\epsfxsize=1.
\columnwidth
\epsfbox{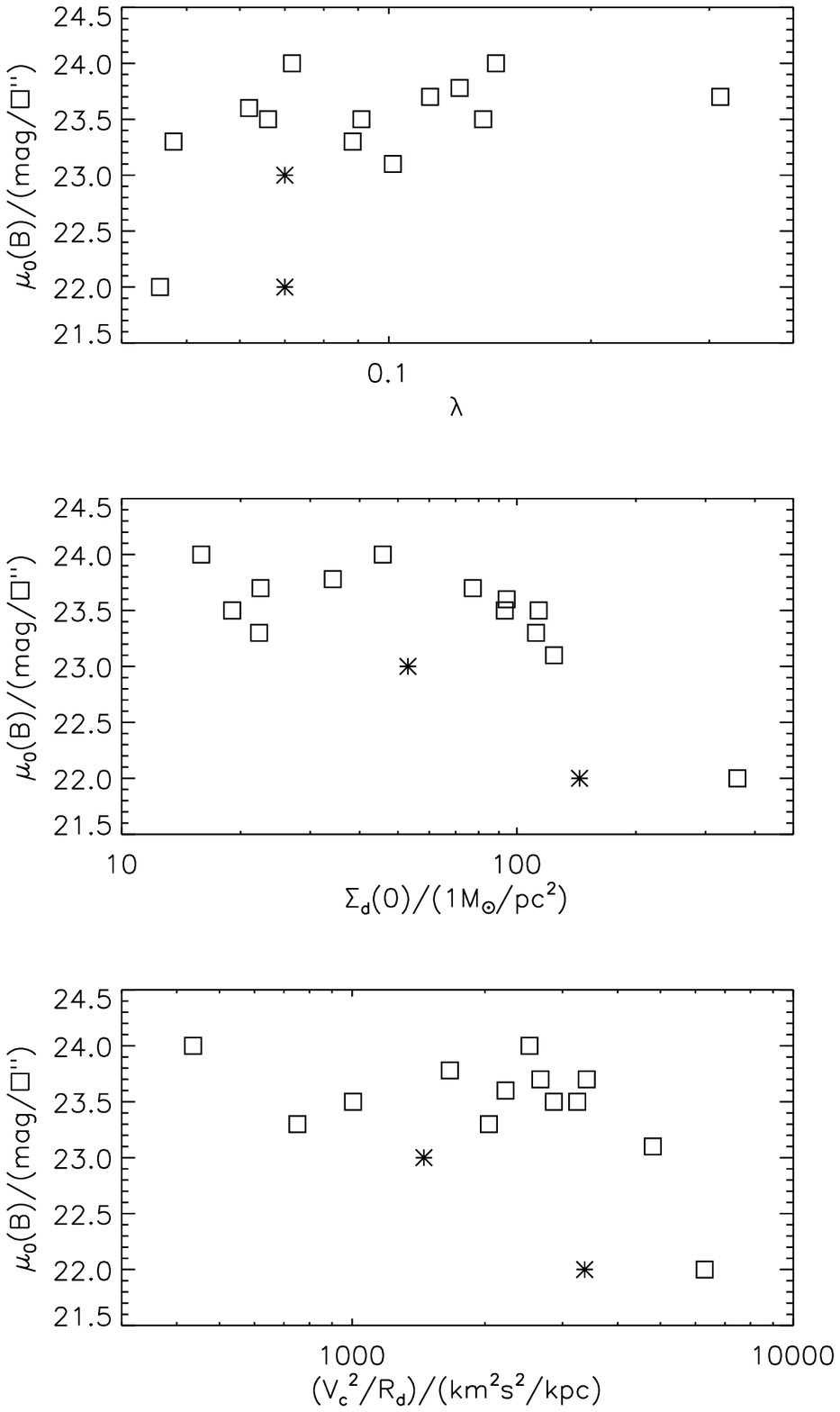}
\caption[]{Upper panel: The central surface brightness is plotted 
versus $\lambda$.

Middle panel: The central surface brightness is plotted versus the central
surface density computed consistently with $M$, $\lambda$, and $z$ of each
galaxy.

Bottom panel: The central surface brightness is plotted versus $V_c^2/R_d$.}
\label{f5}
\end{figure}

\begin{figure*}
\centering
\leavevmode
\epsfxsize=1.5
\columnwidth
\epsfbox{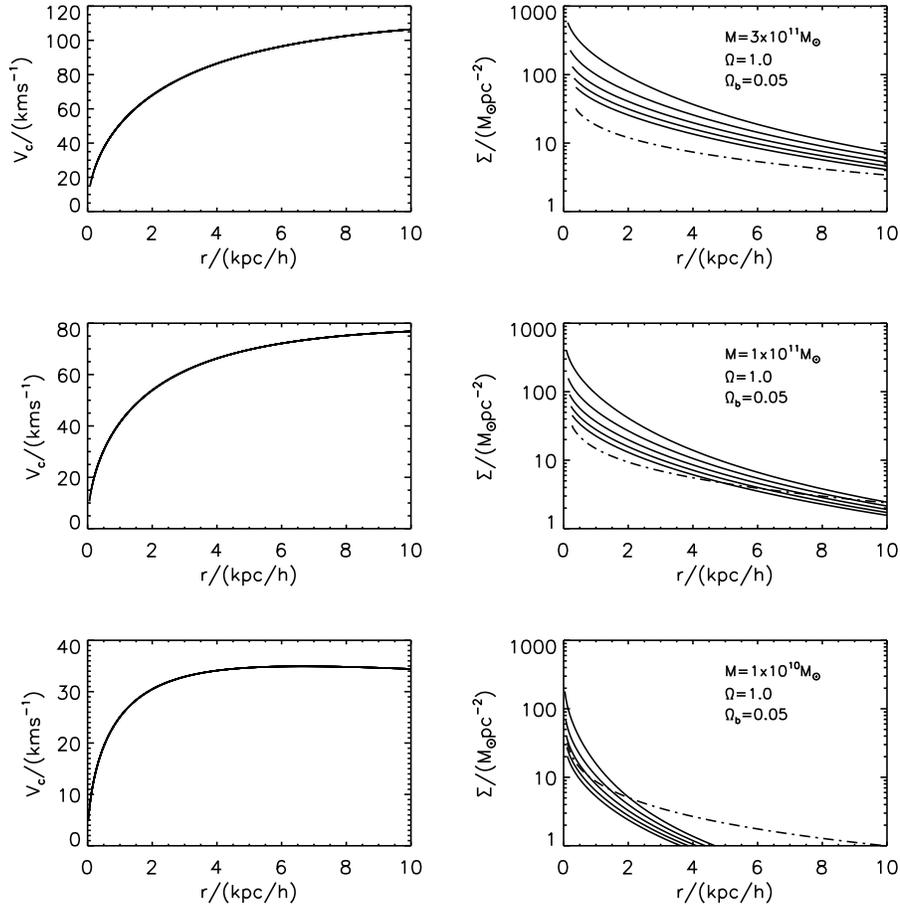}
\caption[]{Rotation curves and density profiles for the JHHP model. The
solid lines correspond to different values for $\lambda$ (0.07 to 0.03, from
top to bottom) and the dashed line is the corresponding critical surface
density to form stars according to the Toomre/Kennicutt criterion. The
predicted initial surface density for the disk is very similar to the
isothermal sphere model.}
\label{f6}
\end{figure*}

A non--self--gravitating disk model is characterized by three parameters:
the halo mass $M$, the spin parameter $\lambda$, and the formation redshift
$z$. For each galaxy in the sample three parameters are known: the 
circular velocity $V_c$, the disk scale--length $R_d$, and $R_1$.
Using these three measured quantities, the unknown parameters $M$, $\lambda$
and $z$ can be determined for each galaxy. Using the equations above one gets:
\begin{equation}
M=\frac{2\pi SR_d^2}{f_b}\exp(R_1/R_d)
\label{10}
\end{equation}   
\begin{equation}
\lambda=\frac{f_bV_c^2}{\sqrt{2}\pi SGR_d}\exp(-R_1/R_d)
\label{11}
\end{equation} 
\begin{equation}
z=\left[\frac{f_bV_c^3\exp(-R_1/R_d)}{20\pi SGR_d^2H_0}\right]^{2/3}-1
\label{12}
\end{equation}
where $S=1$~M$_{\odot}$pc$^{-2}$.  The solution is given in
Table~1 for all galaxies. The observed parameters of F583-1 are
apparently inconsistent with the model, as mentioned above.

\begin{figure*}
\centering
\leavevmode
\epsfxsize=1.5
\columnwidth
\epsfbox{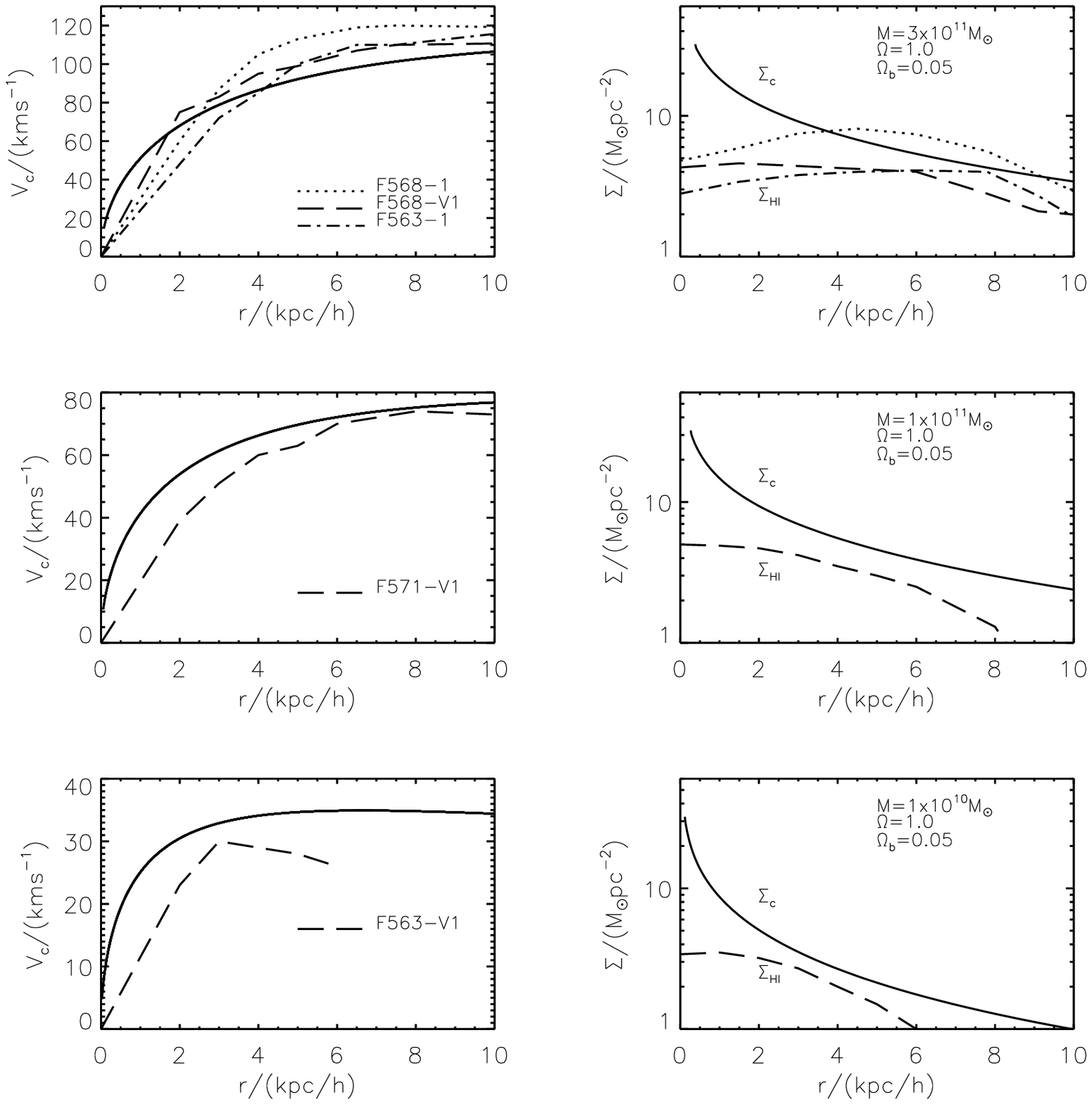}
\caption[]{The model predictions for the JHHP model are compared with a
selected sample of LSBs from \scite{deBlok_phot+95}. The rotation curves are
not well fitted in the inner parts of the halo. The right three panels show
the observed HI surface density (\pcite{deBlok_HI+96}) compared to the
previously computed $\Sigma_c$ using the Toomre/Kennicutt instability
criterion. The agreement between both curves is good beyond 2 Kpc.}
\label{f7}
\end{figure*}

In Fig.~\ref{4} the absolute blue magnitude of the galaxies is plotted versus
the estimated halo mass (from Table~1) and the value of $V_c^2R_d$, that
should be proportional to the halo mass, contained inside the radius $R_d$. In
both plots the absolute magnitude grows with the mass estimator.

In Fig.~\ref{5} the central surface brightness is plotted versus $\lambda$
(upper panel). There is no clear trend in the plot, as could be expected since
the central surface density, and therefore the central surface brightness of a
disk depends on both the spin parameter and the mass.  When the central
surface density of the disk is appropriately computed (equation (7))
consistently with the values of $M$, $\lambda$, and $z$ estimated for each
galaxy (Table~1), a clear correlation between the central surface brightness
and the central surface density can be seen, as shown in Fig.~\ref{5} (middle
panel). The observed central surface brightness grows approximately linearly
with the estimated central surface density, $\Sigma_d(0)$, for
$\Sigma_d(0)>50$~M$_{\odot}$pc$^{-2}$. If the halo formation redshifts and the
spin parameters were unknown, the mass could only be estimated as something
proportional to $V_c^2R_d$, and the central surface density as proportional to
$V_c^2/R_d$. In the bottom panel of Fig.~\ref{5} the central surface
brightness is plotted versus $V_c^2/R_d$. There is no clear trend in the plot,
which shows that a correct estimate of mass and surface density requires a
knowledge of $\lambda$ and $z$ for each galaxy, and not only of its circular
velocity and size. The middle plot of Fig.~\ref{5} is therefore a strong
indication that i) the surface brightness correlates with the surface density,
and ii) the surface brightness is determined by the spin parameter, for any
given halo mass.

\subsection{Non-singular halo}

In \scite{Jimenez+Heavens+97} a model (JHHP model hereafter) was developed for
computing the final settling radius and surface density of the gas of a
rotationally supported disk. In this model it was assumed that the gas
component has the same specific angular momentum as the dark matter, and the
dark matter profile found in simulations by \scite{Navarro+97} was used. A
detailed description of the model can be found in
\scite{Jimenez+Heavens+97}. Using the above we have computed several disk
models for different values of the mass and spin of the halo. Fig.~\ref{f6}
shows the predicted rotation curves and initial surface densities for 3
different masses and 5 values for $\lambda$: 0.03, 0.04, 0.05, 0.06 and 0.07
(from top to bottom). Also plotted is the critical surface density (dashed
line) to form stars according to the \scite{Toomre_64} stability criterion
(see also \scite{Kennicutt_89}). If the surface density of the disk is above
critical, then it will form stars, otherwise it will not.

In Fig.~\ref{f7} we compare the predictions of the JHHP disk model with
observed rotation curves and HI surface densities for F568-1, F568-V1, F563-1,
F571-V1 and F563-V1. The central region of the rotation curves in all cases
does not agree well with the model predictions since the observed central
rotation curves are shallower than the predicted ones. The three right panels
of Fig.~\ref{f7} compare the observed HI surface density with the predicted
critical surface density. In all cases the observed HI surface density and the
predicted critical density agree reasonably well beyond 3-4 kpc showing the
validity of the Toomre criterion, while in the central regions the observed HI
surface density is much lower than the predicted critical density, pointing
towards global instabilities controlling star formation in the central
regions.

The important point to notice here is that the JHHP disk model and the
isothermal sphere predict roughly the same initial surface density for a
rotationally-supported disk (as can be seen by comparing the bottom-right
panel from Fig.~\ref{f6} and the middle panel from Fig.~\ref{f1}). This shows
that the initial surface density is a robust prediction and since, as it will
be shown in the next two sections, it controls what the colours (and therefore
ages) of an LSB will be, we regard our ages and colours as robust predictions.

In order to compare a disk model with the photometric properties of LSB
galaxies, it is necessary to calculate an appropriate stellar population
model, based on the predicted surface density. It is not possible to calculate
models of galaxies without a very detailed and up--to--date treatment of
stellar evolution and atmosphere models, for the simple reason that galaxies
are made of stars. The stellar population model must be based on the
computation of the chemical evolution of the galaxy, consistent with a
reasonable star formation history, because the colours of a stellar population
depend strongly on its chemical composition, and the chemical composition of
the star forming gas is determined by the star formation process itself. In
the next two sections we briefly present the main ingredients of the synthetic
stellar population and of the chemical evolution model. The latter is
basically the same used to reproduce the chemical composition of the Milky
Way, since we are testing the hypothesis that all LSBs originate in the same
way as HSBs, apart from being hosted in dark halos with larger spin parameter.

\begin{figure}
\centering
\leavevmode
\epsfxsize=1.
\columnwidth
\epsfbox{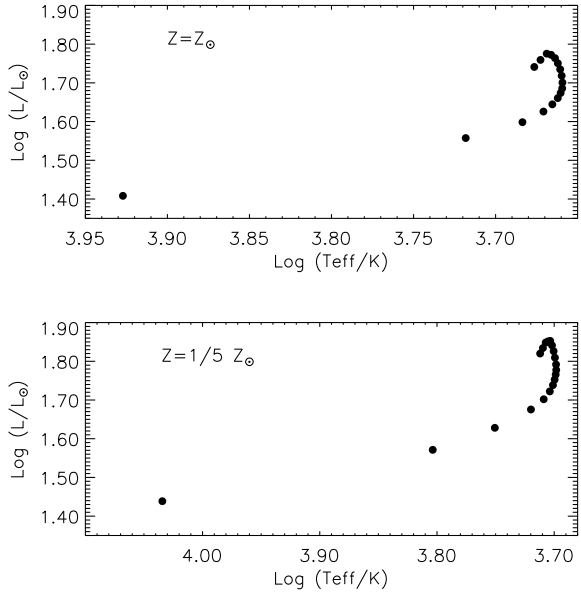}
\caption[]{The Zero Age Horizontal Branch models used in building the
SSPs (see text) for two different metallicities and several masses (0.55 to
1.5 $M_{\odot}$; left to right). Notice the spread in $T_{\rm eff}$ and
luminosity, thus the impossibility to treat the HB like a dot.}
\label{f8}
\end{figure}

\begin{figure}
\centering
\leavevmode
\epsfxsize=1.
\columnwidth
\epsfbox{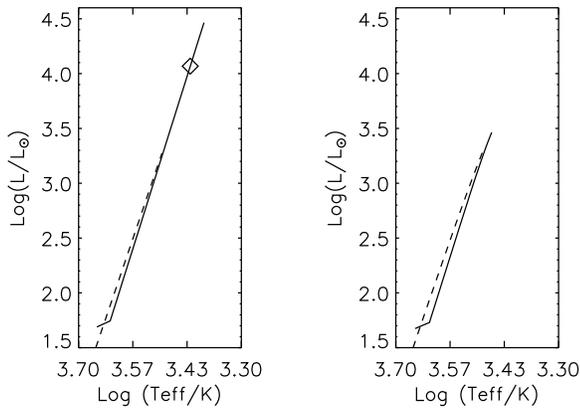}
\caption[]{The evolution along the red giant (dashed line) and asymptotic
branch (solid line) for a 1 $M_{\odot}$ star with solar metallicity with
(right panel) and without mass loss (left panel). In the latter case the star
evolves to higher luminosities and forms a carbon star (diamond) while in the
former case its evolution is stopped at a lower luminosity. Therefore, a
proper modelling of mass loss is important, otherwise the synthetic population
would be too red.}
\label{f9}
\end{figure}
    
\section{Synthetic Spectra and Colours}

In order to compute the spectro-photometric evolution of LSBs we have used a
new set of synthetic stellar population models that are an updated version of
the previous set of \scite{Jimenez+98} models.  Our models are based on the
extensive set of stellar isochrones computed by \scite{Jimenez+98} and the set
of stellar photospheric models computed by \scite{Kurucz_92} and
\scite{Jimenez+98}. The interior models were computed using JMSTAR15
(\pcite{Jimenez+98}) which uses the latest OPAL95 radiative opacities for
temperatures larger than 8000 K, and Alexander's opacities (private
communication) for those below 8000 K. For the stellar photospheres with
temperatures below $8000$ K we have used a set of models computed with an
updated version of the MARCS code (U. J{\o}rgensen, private communication),
where we have included all the relevant molecules that contribute to the
opacity in the photosphere.  For temperatures larger than $8000$ K we have
used the set of photospheric models by \scite{Kurucz_92}.  Stellar tracks were
computed self-consistently, i.e. the corresponding photospheric models were
used as boundary conditions for the interior models.  This procedure has the
advantage that the stellar spectrum is known at any point on the isochrone and
thus the interior of the star is computed more accurately than if a grey
photosphere were used, and therefore we overcome the problem of using first a
set of interior models computed with boundary conditions defined by a grey
atmosphere and then a separate set of stellar atmospheres, either observed or
theoretical, that is assigned to the interior isochrone a posteriori. The
problem is most severe if observed spectra are used because metallicity,
effective temperature and gravity are not accurately known and therefore their
position in the interior isochrone may be completely wrong (in some cases the
error is larger than 1000 K).  A more comprehensive discussion and a detailed
description of the code can be found in \scite{Jimenez+98}.

An important ingredient in our synthetic stellar population models is the
novel treatment of all post-main evolutionary stages that incorporates a
realistic distribution of mass loss. Thus the horizontal branch is an extended
branch (see Fig.\ref{f8}) and not a red clump like in most stellar population
models. Also the evolution along the asymptotic giant branch is done in a way
such that the formation of carbon stars is properly predicted (see
Fig.\ref{9}) and therefore the termination of the asymptotic giant branch. It
is worth noticing that in most synthetic stellar population models a constant
mass loss law is applied resulting in horizontal branches that are simply red
clumps. This leads to synthetic populations that are too red by 0.1 to 0.2 in
e.g. $B-R$ (\pcite{Jimenez+98}).

\begin{figure*}
\centering
\leavevmode
\epsfxsize=1.5
\columnwidth
\epsfbox{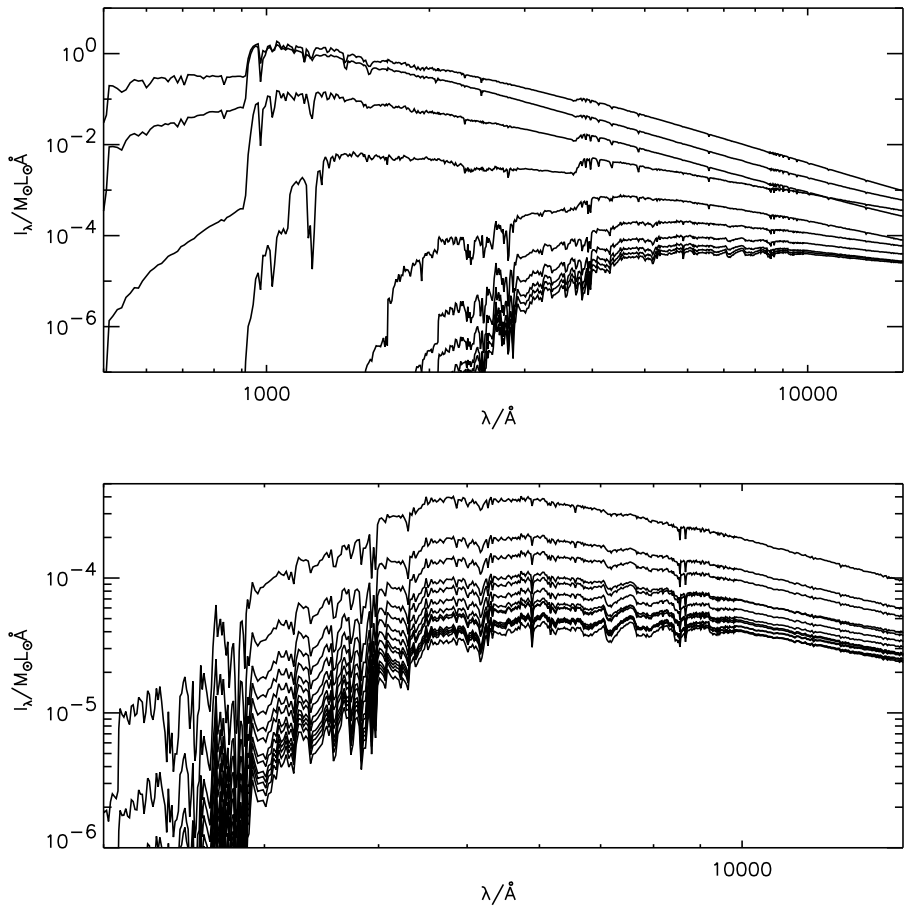}
\caption[]{SSPs for $Z=2Z_{\odot}$. Ages range between $1 \times 10^6$
and $1.5 \times 10^{10}$ yr (from top to bottom).}
\label{f10}
\end{figure*}

Using the above stellar input we compute synthetic stellar population models
that are consistent with their chemical evolution.  The first step is to build
simple stellar populations (SSPs). An SSP is a population of stars formed all
at the same time and with homogeneous metallicity. The procedure to build an
SSP is the following:

\begin{enumerate}
\item A set of stellar tracks of different masses and of the metallicity
of the SSP is selected from our library. 
\item The luminosity, effective temperature and gravity at 
the age of the SSP is extracted from each track. Each set 
of values characterizes a star in the SSP.
\item The corresponding self-consistent photospheric model is assigned to 
each star.
\item The fluxes of all stars are summed up, with weights proportional
to the stellar initial mass function (IMF).
\end{enumerate}

SSPs are the building blocks of any arbitrarily complicated population since
the latter can be computed as a sum of SSPs, once the star formation rate
is provided.  In other words, the luminosity of a stellar population
of age $t_0$ (since the beginning of star formation) can be written as:
\begin{equation}
L_{\lambda}(t_0)=\int_{0}^{t_0} \int_{Z_i}^{Z_f} L_{SSP,\lambda}(Z,t_0-t)\,
dZ\, dt
\end{equation}
where the luminosity of the SSP is:
\begin{equation}
L_{SSP,\lambda}(Z,t_0-t)= \int_{M_{min}}^{M_{max}}SFR(Z,M,t)\,
l_{\lambda}(Z,M,t_0-t)\, dM
\end{equation}
and $l_{\lambda}(Z,M,t_0-t)$ is the luminosity of a star of mass $M$,
metallicity $Z$ and age $t_0-t$, $Z_i$ and $Z_f$ are the initial and final
metallicities, $M_{min}$ and $M_{max}$ are the smallest and largest stellar
mass in the population and $SFR(Z,M,t)$ is the star formation rate at the time
$t$ when the SSP is formed.

We have computed a very large number of SSPs, with ages between $1 \times
10^6$ and $1.5 \times 10^{10}$ years, and metallicities from $Z=0.0002$ to
$Z=0.1$, that is from $0.01$ to $5 Z_{\odot}$.  Fig.~\ref{f10} shows a set of
synthetic spectra of SSPs with $Z=2 Z_{\odot}$ and ages between $1 \times
10^6$ and $1.5 \times 10^{10}$ years (from top to bottom).

Our synthetic stellar population models have been extensively used in previous
works (e.g. \pcite{Dunlop+96,Spinrad+97}), where a detailed
comparison with other models in the literature can be found.

\begin{figure*}
\centering
\leavevmode
\epsfxsize=1.5
\columnwidth
\epsfbox{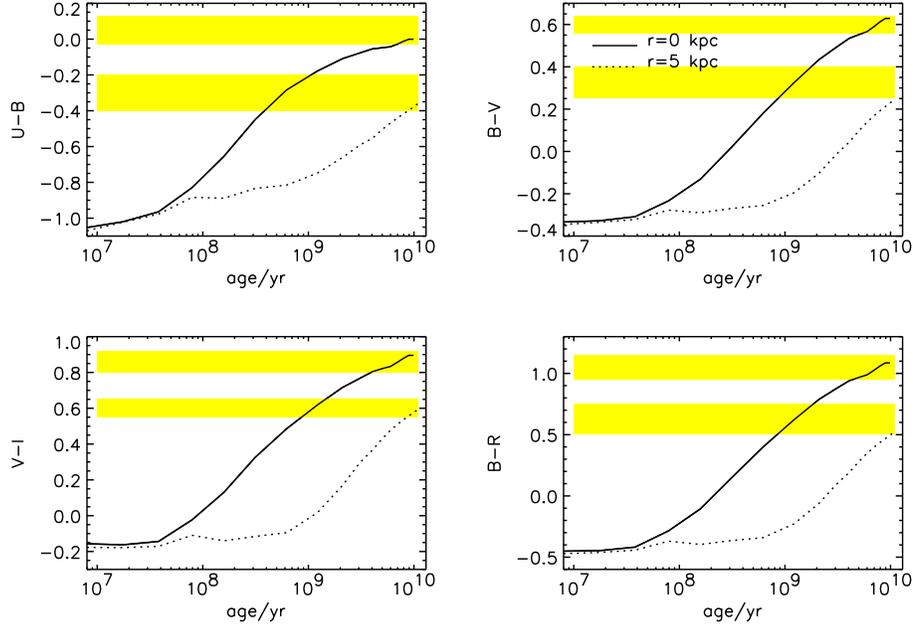}
\caption[]{Time evolution for several colours predicted by our
models. Two radii are shown: the nucleus (solid line) and the outermost radius
of the disk (dotted line). The shaded regions are the observed values for the
bluest galaxies in \scite{deBlok_phot+95} with 1$\sigma$ errors. In all cases
the best fit is at late epochs ($> 7$ Gyr), what shows that LSBs are not young
galaxies.}
\label{f11}
\end{figure*}

\section{Chemical Evolution Models}

Most synthetic stellar population codes do not account for the chemical
time-evolution of the stellar population, and use a simple analytical
prescription for the star formation rate (e.g. the Hubble sequence is often
explained like a convolution of SSPs with constant metallicity (solar) and an
exponentially decaying star formation rate with different e-folding times).
This approach is not self--consistent, but it has been historically used
because of the lack of theoretical libraries of stellar photospheres of
non--solar composition. Indeed, the way to proceed is to use chemical
evolution models where $Z(t)$ and $SFR(t)$ are consistent with each other.
The model of chemical evolution we use is similar to that described in
\scite{Matteucci+89} developed to calculate the chemical evolution of the
Milky Way. The main assumptions of the model are:

\begin{itemize}

\item The disk is represented by several circular concentric shells 2 kpc wide
and no exchange of gas between them is allowed.

\item The disk is formed by infall of primordial material and the infall rate 
is higher in the center than in the outermost regions of the disk.
In particular, the infall law is expressed as:
\begin{equation}
\dot\Sigma_{inf}(r,t)\,=\, A(r) X^i_{inf} e^{-t/\tau(r)}
\end{equation}
where $\tau(r)$ is the timescale for the formation of the disk at a radius 
$r$. The values of $\tau(r)$ are chosen in order
to fit the present time radial distribution of the gas surface density in the
disk. In analogy with what is required for the disk of the Milky Way, we
assumed an ``inside-out'' mechanism of formation of such galaxies, implying
that $\tau(r)$ is increasing towards larger radii (see Table 2).  $X^i_{inf}$
is the abundance of  element $i$ in the infalling gas and the chemical
composition is assumed to be primordial. The parameter $A(r)$ is obtained by
requiring the surface density now to be $\Sigma(r, t_{now})$, 
given by the disk model described in
section 2:
\begin{equation}
A(r)={\Sigma(r, t_{now}) \over \tau (1-e^{-t_{now}/\tau(r)})}
\end{equation}

\item The IMF is taken from \scite{Scalo+86} and is assumed to be constant in 
space and time.

\item The star formation rate is assumed to depend on both the gas surface 
density, $\Sigma_{gas}(r,t)$ and the total mass surface density,
$\Sigma(r,t)$. In particular, we adopted a law of the type:
\begin{equation}
\psi(r,t)= \nu \Sigma_{gas}(r,t)^{k_1} \Sigma(r,t)^{k_2}
\end{equation}
with $k_1=1.5$ and $k_2=0.5$. This choice is the best for the disk of our
galaxy (\pcite{Tosi_82,Matteucci_86}). The parameter $\nu$, which represents
the efficiency of star formation (the inverse of the time scale for star
formation) has been assumed to be $\nu_{lsb}=0.25 Gyr^{-1}$.
\end{itemize}

The effect of stellar winds and outflows is properly accounted for using the
energy output due to supernovae. Therefore, at each ring the model accounts
for the possibility of outflow of gas due to the fact that the temperature of
the interstellar gas is heated enough so the velocity of the gas becomes
larger than the scape velocity. 

The evolution of several chemical species (H, D, He, C, N, O, Ne, Mg, Si, S,
Ca, Fe and Zn), as well as the global metallicity $Z$, is followed by taking
into account detailed nucleosynthesis prescriptions by including the
contributions to galactic chemical enrichment by stars of all masses.  The
instantaneous recycling approximation was relaxed and the stellar lifetimes
were taken into account (for the basic equations see \scite{Matteucci+89}).

\begin{figure*}
\centering
\leavevmode
\epsfxsize=1.5
\columnwidth
\epsfbox{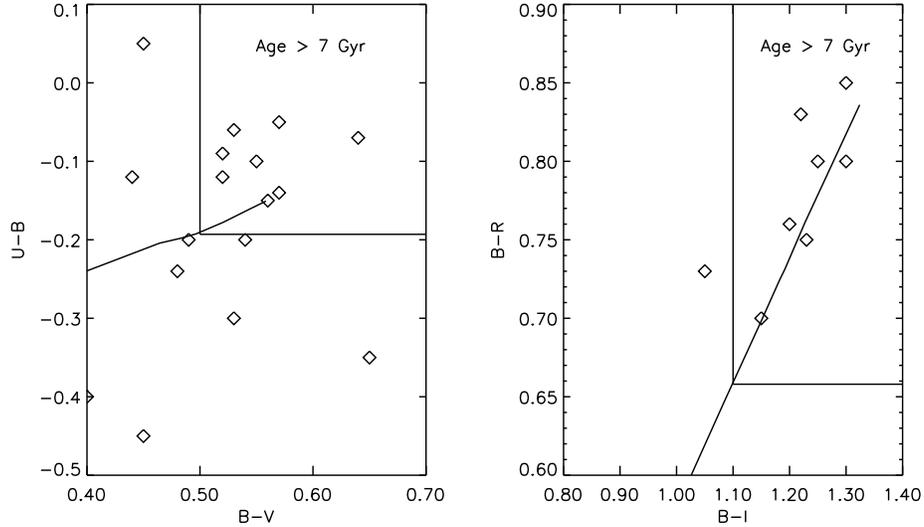}
\caption[]{Area weighted colours for \scite{deBlok_phot+95} sample
compared with our model predictions. Also plotted the region where colours
correspond to ages larger than 7 Gyr. It transpires from the figure that LSBs
have ages in excess of 7 Gyr. Of special interest is the $B-I$ $vs$ $B-R$
plot: it shows a good correlation because it samples the old stellar
population and is not contaminated by recently formed stars like $U$.}
\label{f12}
\end{figure*}

In particular, we adopted the \scite{Renzini+81} stellar yields for low and
intermediate mass stars ($0.8 \le M/M_{\odot} \le 8$) which produce mostly
$^{4}He$, $^{12}C$, $^{13}C$ and $^{14}N$.  Newer stellar yields for this
range of masses are now available (\pcite{Hoek+97}) but they do not differ
significantly from the \scite{Renzini+81} ones (Matteucci et al., in
preparation). For the yields of $\alpha$-elements and Fe in massive stars ($M>
10 M_{\odot}$) we adopted the yields of \scite{Woosley+95} and for type Ia
supernovae (SNe) those of \scite{Thielemann+96}. Type Ia SNe are assumed to be
the outcoming of exploding white dwarfs in binary systems and their
contribution to galactic chemical evolution is computed in the way described
in \scite{Matteucci_Greggio_86}.  They produce mostly Fe and some traces of
$\alpha$-elements.  For the yields of Zn we assumed the prescriptions given in
\scite{Matteucci+93} where this element is assumed to originate partly from
s-processing in massive stars (weak component), partly from r-processing in
low mass stars (main component) and mostly from explosive nucleosynthesis in
SNe of type Ia.

We computed the evolution of a disk with total final mass $M_{d}=
10^{10}M_{\odot}$, $\Omega_b=0.05$, $\Omega_{o}=1$ and $\lambda=0.03$ and 0.06
(see Table 2).

\begin{table}
\begin{center}
\begin{tabular}{ccccc}
$r/kpc$&$\frac{\Sigma}{M_{\odot} pc^{-2}}$&$\tau/Gyr$ &
$Z/Z_{\odot}$&$\frac{\Sigma_{\rm gas}}{M_{\odot} pc^{-2}}$  \\
\hline\hline
0.0    &     210     &     0.05   &     2.70  &   1.8  \\
1.2    &     28      &      1.0   &     1.45  &   2.57 \\
2.5    &     3.86    &      4.0   &     0.60  &   1.2  \\
4.3    &     0.19    &      8.0   &     0.50  &   0.063 \\
\hline\hline
0.0     &      52.68  &      0.05  &    2.35  &    0.79 \\
2.5    &       7.13  &       4.0  &     0.55  &    2.56 \\
4.0     &       2.62  &       8.0  &    0.45  &    0.96 \\
5.0     &       0.13  &       12.0 &    0.30  &    0.04 \\
\hline\hline
\end{tabular}
\caption{Values for the initial surface density ($\Sigma$), $\tau$, final
metallicity ($Z$) and final surface density of the gas $\Sigma_{\rm gas}$ for
the two disc models considered in this work: $\lambda=0.03$ (top panel) and
0.06 (bottom panel).}
\end{center}
\end{table}

\section{The Model Galaxy}

Using the method described in the three previous sections, we have computed
colours and metallicities for the two models described in Table 2.  We have
assumed a halo formation redshift $z=0.7$, a baryon fraction $f_{b}=0.05$, a
density parameter $\Omega_{0}=1.0$ and the Hubble constant
$H_0=75$~kms$^{-1}$Mpc$^{-1}$. As we said before, we use the sample of
\scite{deBlok_phot+95} to compare colors and surface brightness of our models
with the observations. We also compare our predicted oxygen abundances to
those measured in \scite{McGaugh_oxigen94}, and our predicted evolution of Zn
$vs$ redshift with the data by \scite{Pettini+97}.

\subsection{Colour Evolution and Ages}

Fig.~\ref{f11} shows the time evolution of U-B, B-V, V-I and B-R colours for a
model with halo mass $M=1.e10$ M$_{\odot}$, halo redshift formation $z=0.7$,
and spin parameter $\lambda=0.06$. The central radii (solid line) and the
outermost radii (dotted line) are plotted.

Also over-plotted as shaded regions are the mean values of U-B, B-V, V-I and
B-R using F568-1, F571-5, UGC1230, F577-V1, UGC128 and UGC628 from
\scite{deBlok_phot+95} for the central region and for the outermost radii (the
extension of the shaded regions corresponds to the dispersion among the mean
color of the galaxies).  We choose this particular set of galaxies because
they are the bluest in \scite{deBlok_phot+95} sample and therefore they set
the most conservative constraint on the age (lowest age). Two important
conclusions can be drawn from this plot: the best fit occurs in all cases for
ages bigger than 7 Gyr, this happens in the nuclear region as well as in the
outermost radii. Our model provides a good fit for both the nucleus and the
outer radii simultaneously. It is worth mentioning that the observed colours
of the nucleus are rather red, unequivocally indicating the presence of an old
population in the LSBs. This has been recently confirmed by
\scite{Quillen_Pickering_97} who have found a population of old, red stars
($B-H > 3.5$), with colours similar to those of ellipticals, in a sample of
LSBs.

A more interesting constraint on the age can be placed by using colour-colour
plots. In this case we use the area weighted colours for the whole
\scite{deBlok_phot+95} sample to compare our predictions. In Fig.\ref{12} we
show the evolution of our model (area weighted) for two colour plots: $U-B$
$vs$ $B-V$ and $B-R$ $vs$ $B-I$. We have also drawn the lines where the
colours correspond to ages bigger than 7 Gyr. It transpires from the plots
that the observations are better fitted if LSBs are older than 7 Gyr.  The
observed points show a better correlation in $B-R$ $vs$ $B-I$ than in $U-B$
$vs$ $B-V$. This is because $I$ and $R$ sample much better the old stellar
population than $U$. $U$ is affected by recent episodes of star formation, and
this is the possible reason for the scatter in the $U-B$ $vs$ $B-V$
plot. However, there are LSBs with observed $U-B > -0.2$ and $B-V > 0.5$,
which give ages $> 7$ Gyr, in agreement with the $B-R$ $vs$ $B-I$ age
determination.

It is worth noticing that the $\lambda=0.03$ model has colours that are 0.2
redder than the $\lambda=0.06$ model with a similar time evolution. 

These more sophisticated models confirm the results obtained in our previous
paper (\pcite{Padoan+Jimenez+Antonuccio97a}), where we were using the most
simplistic model to show that LSBs have an underlying population older than 7
Gyr.

\subsection{Colour Profiles, Surface Brightness and Spectra}

\begin{figure}
\centering
\leavevmode
\epsfxsize=1.
\columnwidth
\epsfbox{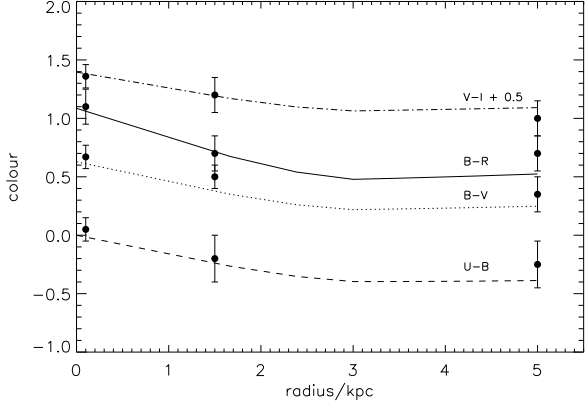}
\caption[]{Predicted colour profiles for the model considered in the text
(lines) are compared with the bluest galaxies in \scite{deBlok_phot+95} sample
(dots, see text). The agreement is excellent but our profiles are slightly
steeper than the observed ones.}

\label{f13}
\end{figure}

An interesting test for our model is to see if it can reproduce the colour
profiles measured by \scite{deBlok_phot+95}. Fig.\ref{f13} shows our model
prediction compared with the mean colours for F568-1, F571-5, UGC1230,
F577-V1, UGC128 and UGC628 from \scite{deBlok_phot+95} sample with error bars
computed as the 1$\sigma$ value of the dispersion among the measured
values. The agreement is in general rather good.  The observed scatter in
\scite{deBlok_phot+95} sample can be easily explained by small changes in the
spin parameter or star formation rate. 

\begin{figure*}
\centering
\leavevmode
\epsfxsize=1.4
\columnwidth
\epsfbox{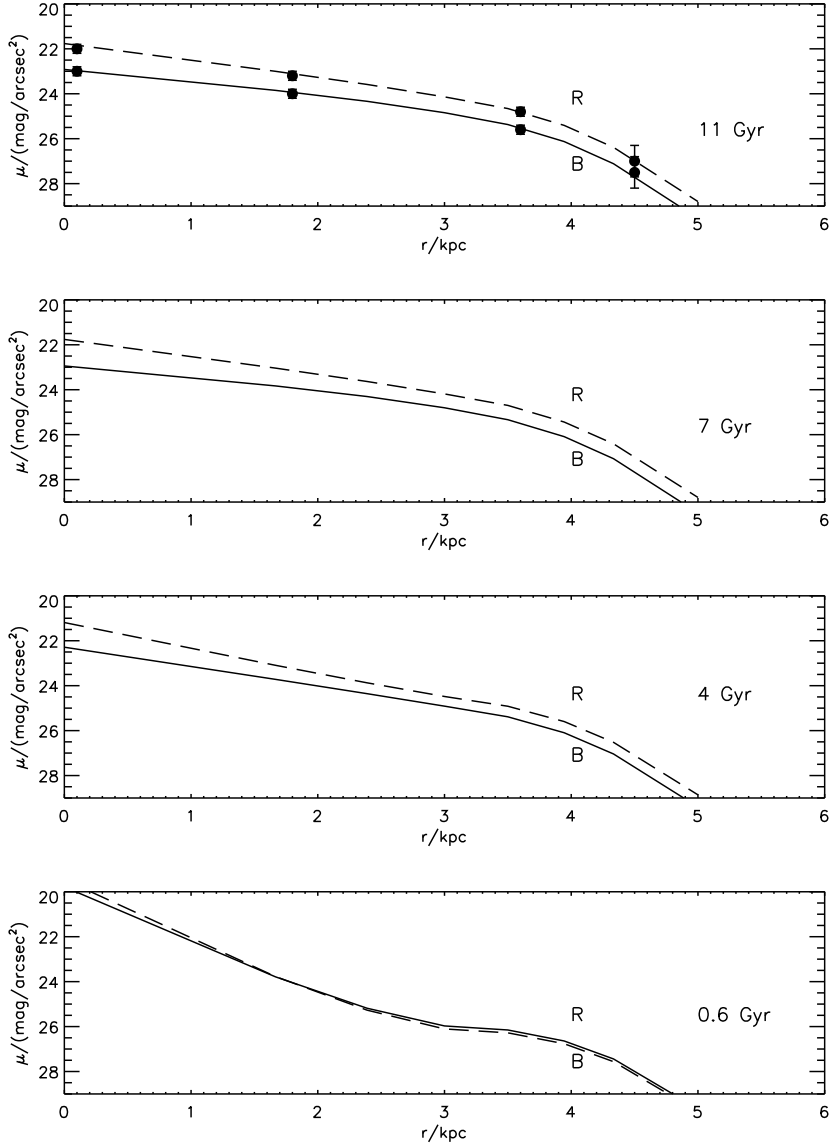}
\caption[]{The surface brightness profiles for our model at different ages are
compared to the corresponding averaged LSB from \scite{deBlok_phot+95} sample
(see text). It is clear that our model gives a good fit both for the profile
and the absolute value of the surface brightness in both bands. It also
transpires from the figure that ages as low as 4 Gyr do not provide a good
fit, reinforcing the old age argument presented in this work.}

\label{f16}
\end{figure*}

We have also computed the radial surface brightness for different times
(Fig.\ref{f16}). We compare it with the averaged values for F571-5, UGC1230,
F577-V1 and again the error bars correspond to the 1$\sigma$ value of the
dispersion among the measured values.  Our model is quite successful in
fitting the observed surface brightness gradients. There is
little evolution with time in the surface brightness after 7 Gyr, while
the model is definitely brighter than the observed galaxies, for ages
smaller than 5 Gyr.

\begin{figure*}
\centering
\leavevmode
\epsfxsize=1.7
\columnwidth
\epsfbox{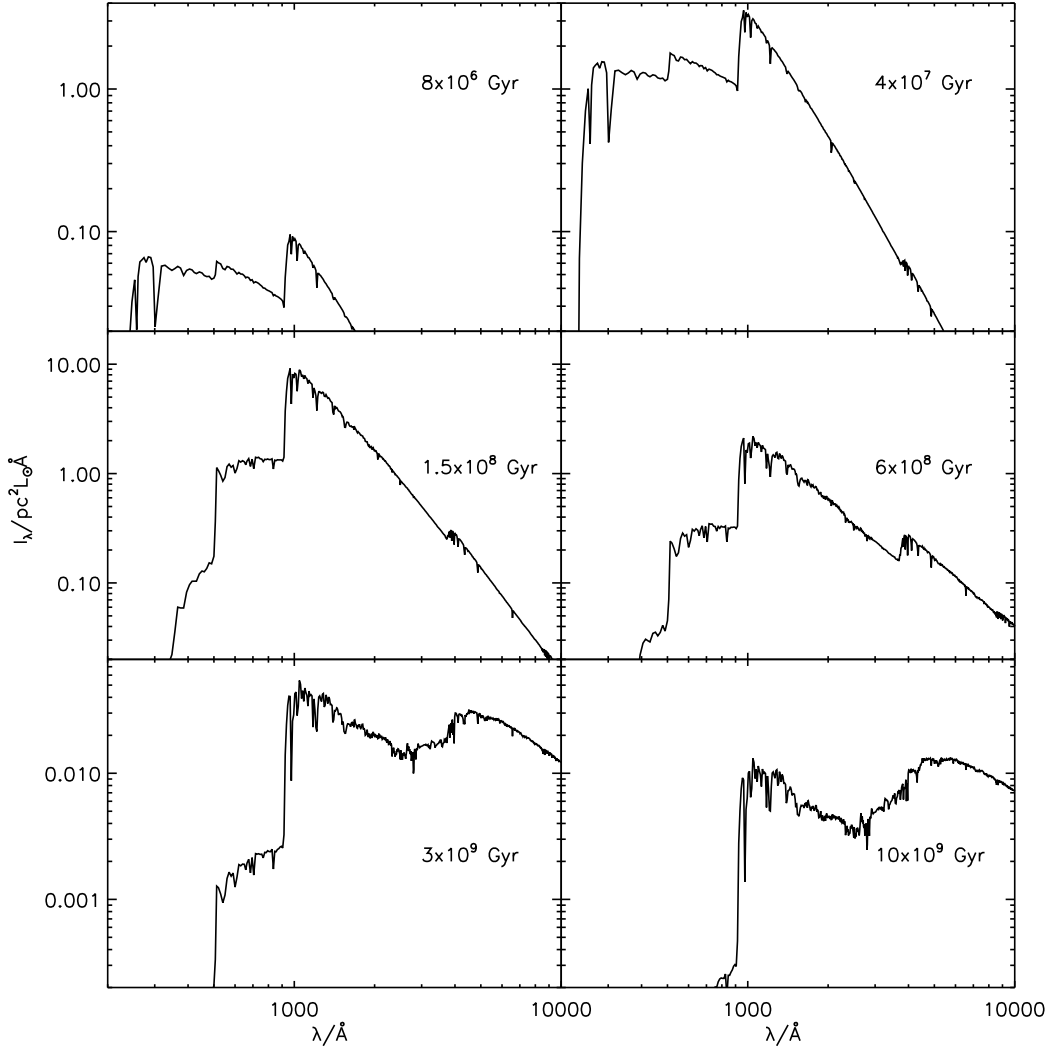}
\caption[]{Predicted spectra for the nucleus of a typical LSB.}
\label{f14}
\end{figure*}

\begin{figure*}
\centering
\leavevmode
\epsfxsize=1.7
\columnwidth
\epsfbox{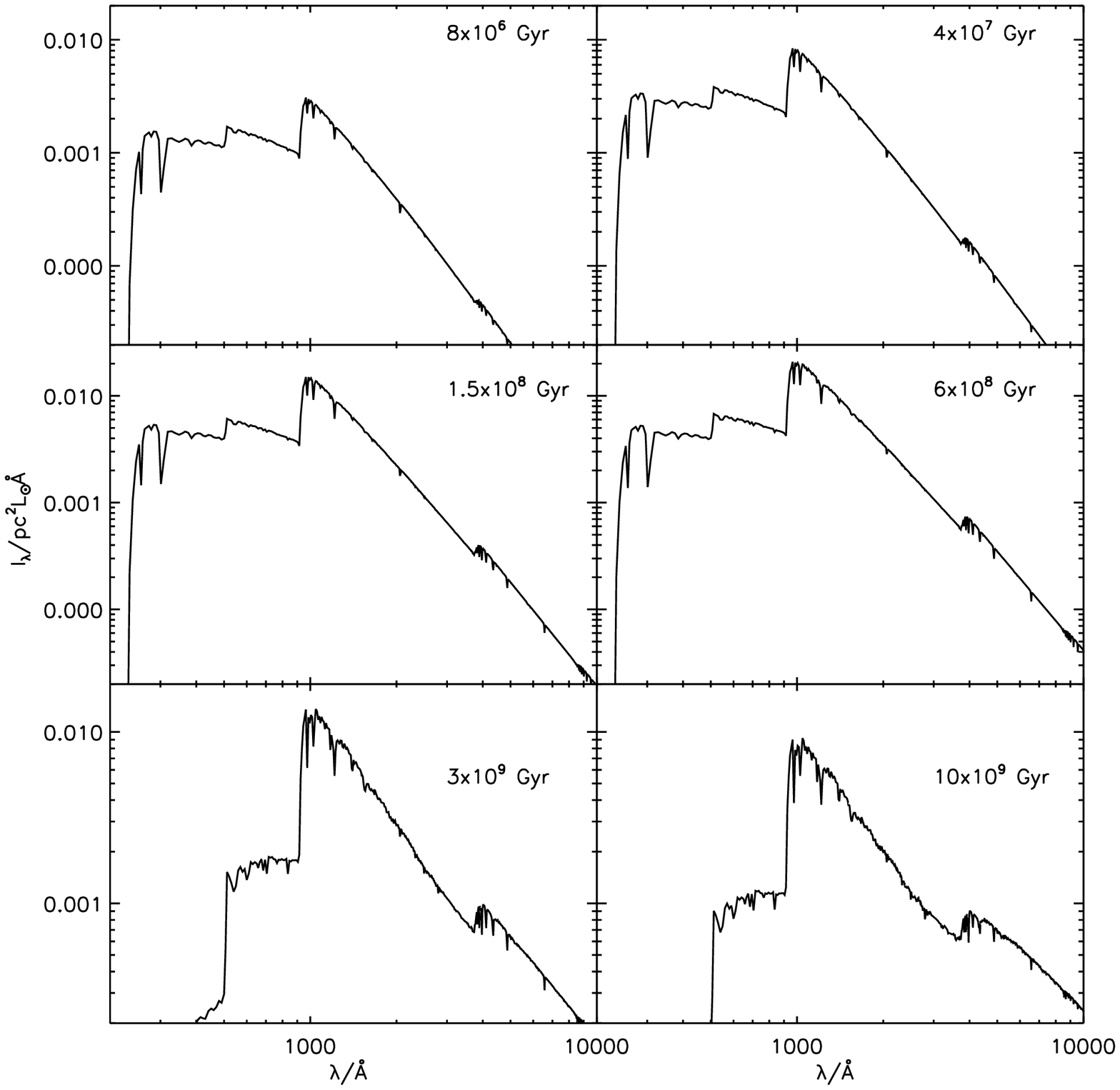}
\caption[]{Predicted spectra for the outermost radius of a typical LSB.}
\label{f15}
\end{figure*}

Fig.\ref{f14} and \ref{f15} show the predicted spectra for six different ages
in the nuclear region (Fig.\ref{f14}) and in the outermost radius
(Fig.\ref{f15}).

\subsection{Chemical Abundances} 

In the previous sections we have shown that the first stars in LSBs formed
about 8-9 Gyr ago.  One observable which can potentially cast doubt on the
model is the metal abundance, for example the Zinc measurements of
\scite{Pettini+97}.  In order to ensure that the metal abundance is not
overpredicted, we present an illustrative computation which is designed to
maximise the predicted abundance.  We assume a halo collapse at $z=4$,
corresponding to a look-back time of 8 Gyr if $H_0=75$ km s$^{-1}$ Mpc$^{-1}$.
This extreme collapse redshift would correspond to rare high-peak fluctuations
(e.g. \pcite{Peacock_Heavens_85}), and leads to high surface densities, high
infall rates and, of course, an early build-up of metals.  It should therefore
be regarded as providing a practical upper limit to the elemental abundances
computed in this section.

\begin{figure*}
\centering
\leavevmode
\epsfxsize=1.4
\columnwidth
\epsfbox{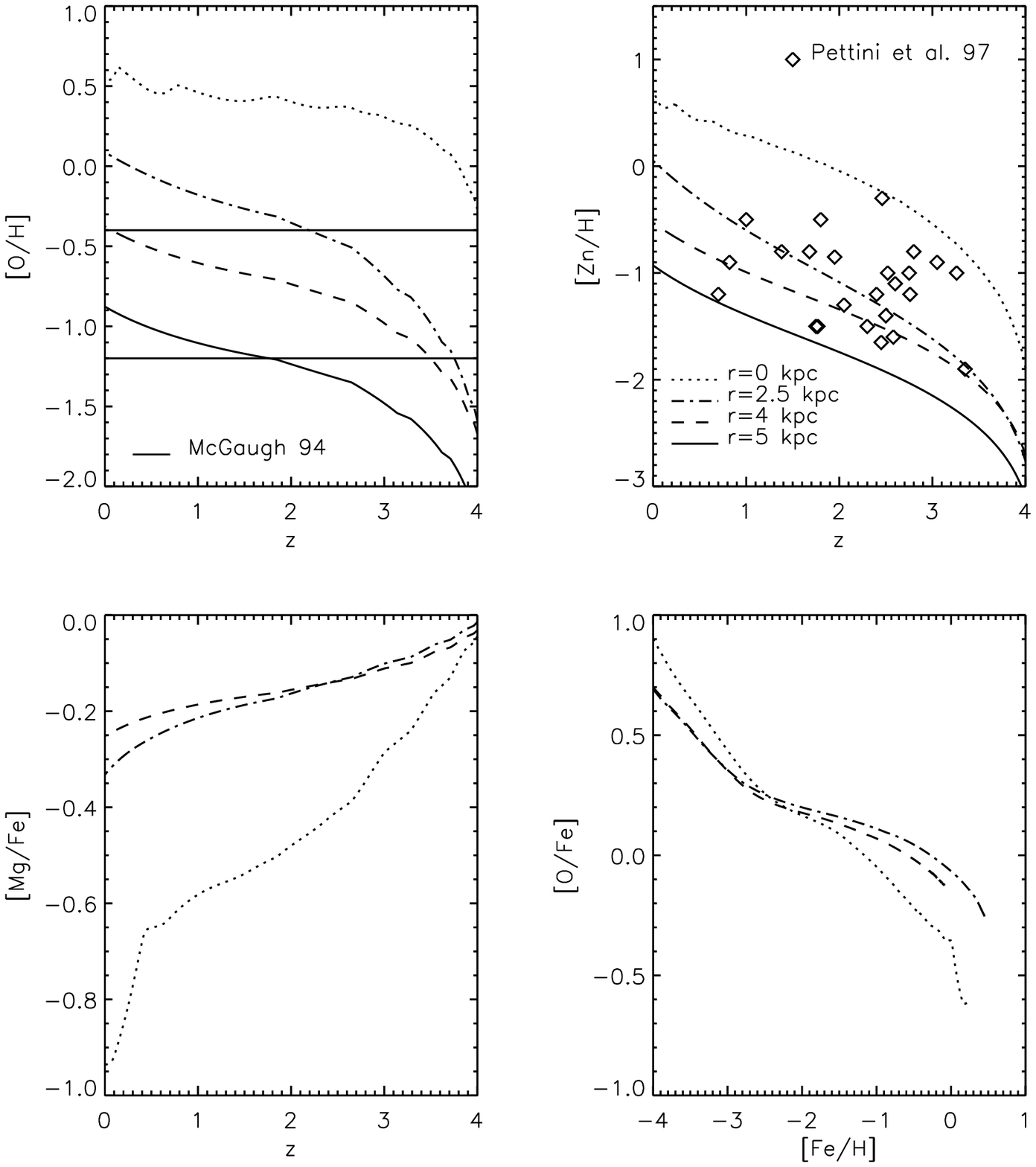}
\caption[]{The redshift evolution for some elements is plotted. The
top-left panel shows the evolution of $[O/H]$ and is compared with
\scite{McGaugh_oxigen94} measures. Our model reproduces perfectly the peak in
$[O/H]$ among LSBs found by \scite{McGaugh_oxigen94} at $[O/H]=-0.9$. The
top-right panel shows the analogous to the previous one but for $[Zn/H]$
compared with \scite{Pettini+97} measures. It transpires from the figure that
the predicted $[Zn/H]$ for LSBs is consistent with the spread found for DLAs.}
\label{f17}
\end{figure*}

Using our chemical evolution model it is possible to predict the evolution
with radius and redshift for most chemical species. In this section we compare
the prediction of our model with two samples.  \scite{McGaugh_oxigen94} has
observed HII regions in a sample of LSBs and derived their oxygen
abundance. Most of the HII regions are located in the outermost part of the
spiral arms. He finds quite a spread in oxygen abundance among LSBs, with a
clear peak at about $[O/H]=-0.9$ and dispersion 0.15 (FWHM). His measures also
show a tentative peak at about $[O/H]=-0.4$, but as is said in
\scite{McGaugh_oxigen94}, it is partly an artifact of the method employed to
determine the oxygen abundances and not a definitive feature of bimodality. In
Fig.\ref{f17} (upper-left panel) we show our model predictions for four radii
and also show the range of oxygen abundances measured by
\scite{McGaugh_oxigen94} as thick horizontal solid lines.  It is clear that
our model is consistent with the \scite{McGaugh_oxigen94} measures, but does
not explain the lowest abundances measured by \scite{McGaugh_oxigen94}. This
may be explained by the fact that the outermost radius is always close to (or
below) the Toomre/Kennicutt threshold and it is therefore likely that some HII
regions will be forming stars in areas of the galaxy with metallicity close to
primordial values, while the average metallicity of the galaxy at that radius
(that is what we compute) will be higher. In fact, this is well supported by
the fact that the colour profiles from the sample by \scite{deBlok_phot+95}
always show redder colours in the outermost radius than the ones corresponding
to a galaxy with $[O/H]$ as low as $-1.2$. For the outermost radius our model
predicts $[O/H]=-0.8$ in excellent agreement with the peak found by
\scite{McGaugh_oxigen94}.  It should be also noted that none of the innermost
radii in our model reproduces \scite{McGaugh_oxigen94} measures, in good
agreement with the well known fact that HII regions are usually in the
outermost parts of the hosting galaxy.

Another important prediction of our model is the evolution of Zn. The
right-upper panel of Fig.\ref{f18} shows the predicted evolution of Zn for
four different radii in our model. We also show as diamonds the data for
Damped Lyman-$\alpha$ (DLA) systems observed by \scite{Pettini+97}, including
some upper limits. While it is difficult to establish a correlation of Zn $vs$
redshift, it transpires from the figure that the predicted Zn abundance for
LSBs at different radii is {\em consistent} with the \scite{Pettini+97}
data. In fact the spread found at every redshift is consistent with the spread
in Zn abundance predicted for LSBs. It should be noted though, that this does
not mean that DLAs are LSBs, this only means that due to their old age and
their presence at high-$z$, some DLAs could be caused by LSBs in the line of
sight. Indeed, at least one DLA has been identified as a LSB galaxy
(\pcite{Pettini+94}).

Notice that the results derived in this section are robust for $z=0$
regardless of the formation redshift adopted for LSBs since the evolution of
the chemical elements with time is very weak after the first 4 Gyr.

\section{Discussion and Conclusions}

\subsection{The Nature of LSBs}

In this work we have shown that the photometric properties of the bluest
galaxies in the sample by \scite{deBlok_phot+95} can be reproduced assuming
that they form and evolve as normal disk galaxies, with relatively high spin
parameter. That the spin parameter alone could explain the low surface
brightness was already shown by \scite{Dalcanton_disc+97}, but they assumed a
given $M/L$ ratio. Here we have strengthened the point by also explaining the
colours, the colour gradients, and the chemical abundances. We have also
provided synthetic spectra that could be compared with the observations, in
the effort to prove that LSB disk galaxies are indeed normal galaxies with
high spin parameter.  The stellar populations of LSBs are rather evolved,
especially in the central part of their disks. Even the bluest galaxies in the
LSB sample by \scite{deBlok_phot+95} are older than 7~Gyr, and typically star
formation in these galaxies starts about 9~Gyr ago. This confirms the results
of \scite{Padoan+Jimenez+Antonuccio97a}, where the galaxies in the same sample
were found to be at least 7~Gyr old.

LSB and HSB disks are therefore hosted in similar dark matter halos that
differ only for their spin parameter, and they evolve in a similar way, apart
from the fact that LSB disks are less concentrated than HSB disks. This means
that the present day (or low redshift) abundance of LSBs relative to HSBs
should be about the same at any redshift. Moreover, the conclusions about the
epoch of star formation and halo formation (see the next section), and about
the colour evolution must be valid for disk galaxies in general.  In
particular, Fig.~\ref{f11} shows that the colour evolution of LSBs is very
strong. The U-B, B-V and V-I colours of the central part of the disk at high
redshift are about 1~mag bluer than they are at low redshift, and the B-R
colour even 1.5~mag bluer.  This must be a property of disks in general, and
not only of LSB disks.

\subsection{Star Formation and Galaxy Formation}

We have shown in section 2.1 that the size of LSB disk galaxies is such that
their halos are formed in the redshift range $0.1<z<1.2$. We have also been
able to estimate the formation redshift of the halo of each galaxy in the
sample, and we have found a mean value $z=0.7$.  In an Einstein--de--Sitter
universe with $H_0=75$~kms$^{-1}$Mpc$^{-1}$, a redshift $z=0.7$ corresponds to
a look--back time of $6.8$~Gyr.  On the other hand, the photometric properties
of LSB galaxies, in particular their colours, are such that star formation
must have started approximately $9$~Gyr. Therefore, the star formation starts
about 2~Gyr before the galactic halos are formed. These numbers are likely to
be similar for HSB disk galaxies too, since they have colours and sizes very
similar to LSBs, also suggesting a rather low formation redshift for
their halos.

The discrepancy between the epoch when star formation starts in a galaxy and
the epoch when the halo forms means that the process of star formation starts
before the galactic halo has assembled, as predicted for example by bottom--up
galaxy formation scenarios, characterized by power spectra with more power on
sub--galactic scales than on galactic ones.

\begin{figure}
\centering
\leavevmode
\epsfxsize=1.0
\columnwidth
\epsfbox{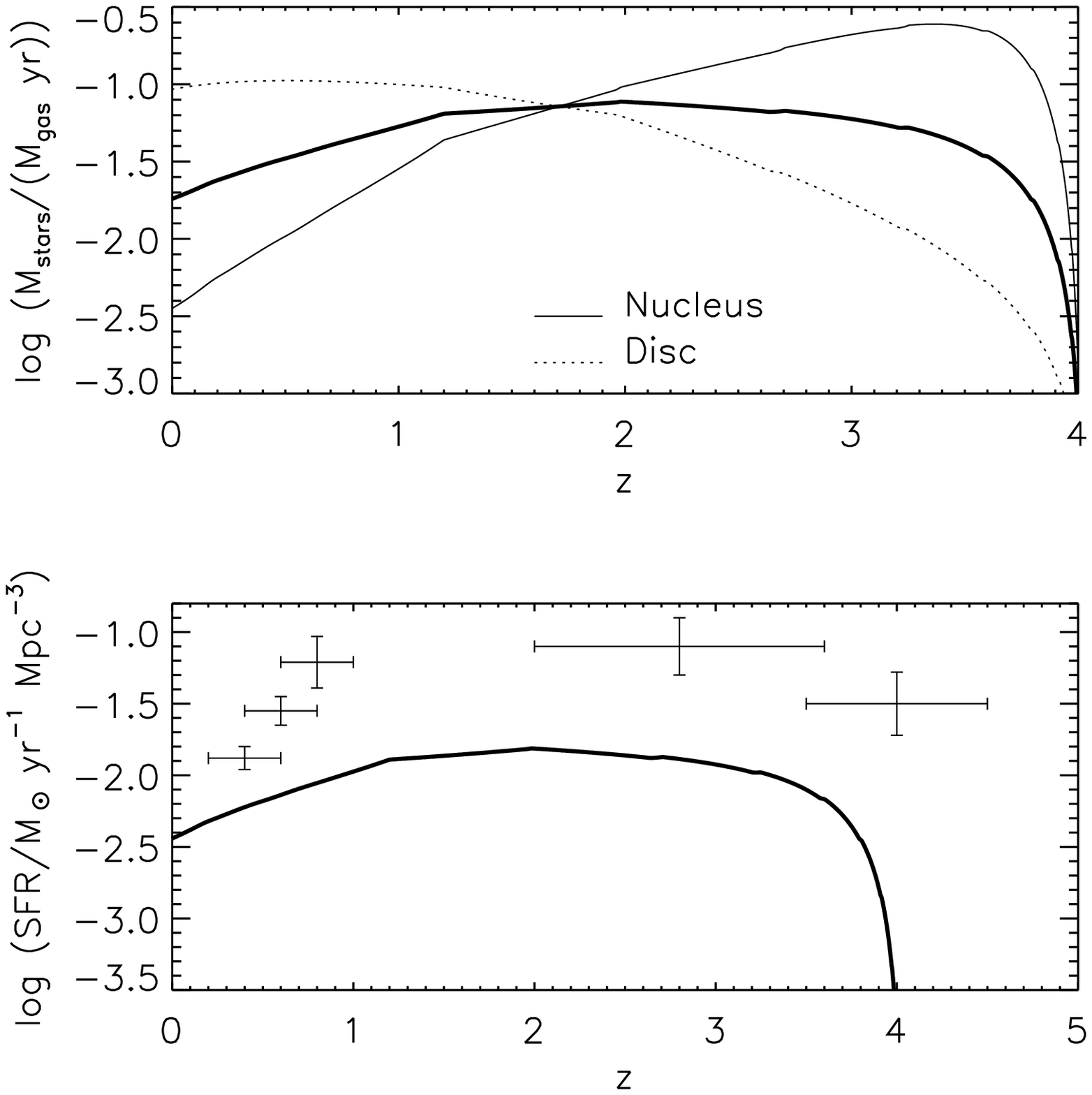}
\caption[]{Top panel shows the gas fraction with time in our LSB model
that is converted into stars $vs$ redshift for the disk and the nucleus. The
thick line corresponds to the average of the nucleus and the disk and should
be a fair representative of the whole LSB. The bottom panel shows the
predicted global SFR in the Universe due to LSBs assuming that the mass for a
$L_*$ LSB is 5 smaller than the mass for a $L_*$ HSB and that the comoving
number density for LSBs is the same as for HSBs (see text). The observed
points are taken from \scite{Madau_97}. Notice that the evolution with
redshift is flatter than the one measured by \scite{Madau_97}.}
\label{f18}
\end{figure}

In Fig.~\ref{f18} we show the evolution with redshift of the SFR. The upper
panel shows the fraction of gas with time that is converted into stars at
every redshift for the nucleus and the disk. The continuous thick line shows
the average of the nucleus and the disk as a representative of the whole
LSB. It is worth noticing that most of the gas in the nucleus is converted
into stars between $2 < z < 4$, while most of the star formation in the disk
takes place at $ z < 2$. The lower panel shows the global SFR in the Universe
for LSBs assuming that they have the same comoving number density as HSBs and
that the typical mass for a $L_*$ LSB is a factor 5 smaller than the typical
mass of a normal $L_*$ galaxy. The data points are taken from \scite{Madau_97}
and trace the global SFR in the Universe for HSBs. LSBs lie below the observed
points because they have typical masses smaller than HSBs. The important
point to notice here is that the global SFR for LSBs has a flatter evolution
than the observed points.

\subsection{Discussion}

In order to explain the rotational properties of LSBs, we require halos with
velocity dispersions $\approx 100$ km s$^{-1}$, or masses $\approx 10^{11}$
M$_\odot$.  This is also consistent with the required gas masses and a baryon
fraction of a few percent.  To obtain the right sizes of LSBs, we additionally
require a relatively late halo collapse, $z \simeq 0.7$.  Clearly this may
present a problem for galaxy formation models which have a lot of power on the
appropriate scale.  An obvious example is COBE-normalised CDM, with shape
parameter $\Gamma=0.5$ \cite{FW91}, for which the characteristic collapse
redshift for $100$ km s$^{-1}$ halos is around $z=6$.  Reducing the amplitude
to that of standard CDM (bias parameter 2.5) alleviates the problem to a
certain extent, but the collapse redshift is still $z=3$.  It is more
promising to consider mixed dark matter models, for which the power spectrum
is in any case a better match to the observed galaxy spectrum.  For a neutrino
fraction of 30\%, the comoving number density of halos with mass exceeding
$1.5 \times 10^{11}$ M$_\odot$ reaches 0.01 $h^3$ Mpc$^{-3}$, comparable to
the density of large galaxies, at a redshift of about 1.7 \scite{KBHP95}.
This is still a little high, but it should be remembered that the collapse of
halos of given mass does take place over a reasonably wide range of redshifts,
and some fraction of halos would form after $z=0.7$.  In the model considered
by \scite{KBHP95}, about a quarter of halos exceeding this mass form after
$z=0.7$.  Analytically, there is expected to be a weak correlation between
high spin and late collapse \scite{HP88}, and numerical simulations by
\scite{Ueda94} show an anticorrelation between spin and density.

We postulate therefore that the LSBs form in halos with high spin which form
relatively late, in the context of a hierarchical model.  The indications are
that a model with relatively little small-scale power, such as mixed dark
matter, would do best here, but there are enough uncertainties in the collapse
redshift, both in the simplifications of the halo model, and in physical
processes which might delay collapse (e.g. \pcite{BabulRees92}) that this
conclusion is not strong.

In summary, the most important results of this work are:

\begin{enumerate}
\item Observed colour profiles, chemical abundances and surface brightness
profiles for LSBs are well fitted if they are assumed to have a spin parameter
for its halo higher than HSBs.
\item LSBs are not young objects, as often claimed in the literature, since
their colours are well fitted by old ($> 7$ Gyr) stellar populations.
\item There is discrepancy between the photometric age of the galaxies and the
age of formation of their halo, which indicates that star formation can start
about 2~Gyr before the halo is formed.  This is perfectly acceptable in
hierarchical models of galaxy formation, and, as discussed in the text, the
discrepancy can be reduced or removed if the cosmological model we have
assumed is incorrect. If an open ($\Lambda$) cosmology had been adopted, the
above discrepancy would have been completely removed.
\end{enumerate}

\section*{acknowledgements}
We thank the anonymous referee for helpful comments that improved this paper
and Max Pettini for a careful reading of the manuscript and helpful
discussions on the nature of DLAs. We also thank Johannes Blom for a careful
reading of the manuscript.

\end{document}